\documentclass[a4paper]{amsart}
\usepackage[english]{babel}
\usepackage[latin1]{inputenc}
\usepackage{amsmath}
\usepackage{amsthm}
\usepackage{graphicx}
\usepackage{rotating}



\newcommand{\snitt}[1]{[#1, \bar{#1}]}
\newcommand{\dd}{\mathrm{d}}

\newcommand{\magn}{\mathcal{M}}
\newcommand{\enrg}{\mathcal{E}}

\theoremstyle{plain}
\newtheorem{theorem}{Theorem}
\newtheorem{corollary}[theorem]{Corollary}
\theoremstyle{definition}
\newtheorem{definition}[theorem]{Definition}

\title{Series expansion for the density of states of the Ising and
  Potts models}
\author{Daniel Andr\'en}

\begin{document}
\maketitle


\section{Introduction}
The Lentz-Ising model of ferromagnetism has been thoroughly studied
since its conception in the 1920's\cite{lenz}.  It was solved in the
1-dimensional case by Ising himself in 1925\cite{ising} and in the
2-dimensional case without an external field by Onsager in
1944\cite{onsager}.  For an introduction to the model, see Cipra
\cite{cipra}.
    
The partition function on a graph \(G\) on \(n\) vertices and \(m\)
edges is defined as
\begin{equation}
	Z(G;x,y)=\sum_{i,j} a_{ij} x^i y^j.
\end{equation}
Here \(a_{ij}\) counts the number of induced subgraphs of \(G\) with
\((n-j)/2\) vertices and \((m-i)/2\) edges in the boundary.  We refer
to the index \(i\) as the energy and the index \(j\) as the
magnetization.

The traditional partition function studied in statistical physics is
then obtained by evaluating it at a certain point
\begin{equation}
	Z \left( G; e^K, e^H \right) 
\end{equation} 
where \(K = -J/k_{B}T\), \(H = -h/k_{B}T\) and \(J\) and \(h\) are
parameters describing the interaction through edges and with an
external magnetic field respectively, \(T\) is the temperature and
\(k_{B}\) the Boltzmann constant.


The main goal in the study of the Ising model on a graph \(G\), or
some family of graphs, is usually to study the model in the vicinity of
a \emph{critical temperature}, denoted \(T_{c}\), where the model
undergoes a phase transition and there determine the behaviour of
various critical properties.



\section{Definitions and notation}
Let \(G=(V,E)\) be a graph with vertex set \(V\) with \(|V|=n\)
vertices and edge set \(E\) with \(|E|=m\) edges. Let a state
\(\sigma\) be a function from the vertices \(V\) to the set \(\{\pm
1\}\) and let \(\Omega\) be the set of all states. We can then define
the \emph{energy} of the graph \(G=(V,E)\) in state \(\sigma\) to be
\begin{equation}
  \enrg(G,\sigma) = \sum_{uv \in E} \sigma(u) \sigma(v)
\end{equation}
the \emph{magnetization} to be
\begin{equation}
  \magn(G, \sigma) = \sum_{v \in V} \sigma(v)
\end{equation}
and formulate a generating function that counts them all as
\begin{equation}
  Z(G; x, y) = \sum_{\sigma \in \Omega} x^{\enrg(G,\sigma)}
  y^{\magn(G,\sigma)} = \sum_{ij} a_{ij} x^i y^j
\end{equation}
where the last equality defines the coefficients \(a_{ij} =
a_{ij}(G)\).  We often drop \(G\) when we can deduce the graph from
the context.

We will also need these definitions later:
\begin{definition}[T-join]
  A T-join \((T,A)\) in a graph \(G=(V,E)\) is a subset \( T \subseteq
  V \) of vertices and a subset \( A \subseteq E \) of edges such that
  each vertex in \(T\) is incident with an odd number of edges in
  \(A\) and each vertex in \(V \setminus T\) is incident with an
  even number of edges from \(A\).
\end{definition}
Observe that the cardinality of \(T\) has to be even since we can not
have a subgraph with an odd number of vertices of odd degree.
\begin{definition}[Cut]
  A cut \(\snitt{S}\) in a graph \(G=(V,E)\) is a subset of edges,
  induced by a partition \(S \cup \bar{S} = V\), that have one
  endpoint in \(S\) and the other in \(\bar{S}\). Let \(|\snitt{S}|\)
  be the number of edges in the cut.
\end{definition}

\begin{definition}[Locally vertex transitive]
  We say that a sequence of graphs \(\{G_i\}_{i=1}^{\infty}\) are
  locally vertex transitive if for each \(R\) there exists an \(N\)
  such that all balls of radius \(R\), in the graph metric, around
  each vertex in each graph in the subsequence
  \(\{G_i\}_{i=N}^\infty\) are isomorphic.
\end{definition}


\section{Series expansion}
We can try to make our problem simpler by setting \(y=1\) and get
\begin{equation}
  Z(x,1) = Z(x) = \sum_{ij} a_{ij} x^i 1^j = \sum_i a_i x^i
\end{equation}
where once again the last equality is the definition of the \(a_i\)
coefficients. What do these coefficients count? Since the state
\(\sigma\) partitions the vertex set \(V\) in two parts and the
\(\enrg(G,\sigma)\) counts edges with one vertex in one part and the
other vertex in the other part as negative and the other edges as
positive we get that \(a_i\) is twice the number of cuts of size
\(\tfrac{m-i}{2}\) (we count each cut twice since we can interchange
the partitions). This has a natural reformulation using even
subgraphs, namely:
\begin{theorem}[van der Waarden]\label{th:vdW}
  Let \(a_i\) be the number of cuts of size \(\tfrac{m-i}{2}\) and let
  \(b_i\) be the number of even subgraphs with \(i\) edges, then:
  \begin{equation} \label{vdW}
    \sum_i a_i e^{iK} = 2 \cosh^m K \sum_i b_i \tanh^i K
  \end{equation}
  \begin{proof}
    The first sum in (\ref{vdW}) is the moment generating function for
    the sequence \(a_i\). The \(k^{\rm th}\) moment of \(a_i\) can be
    written as
    \begin{equation}
      \mu_k = \sum_i a_i i^k = \sum_{\sigma \in \Omega} \left(
        \sum_{uv \in E} \sigma(u) \sigma(v) \right)^k
    \end{equation}
    We now expand the multinomial \( \left( \sum_{uv \in E} \sigma(u)
      \sigma(v) \right)^k \) where each term can be seen as an choice
      of \(k\) out of \(m\) edges (not necessarily distinct).  Now
      observe that if we have chosen an even number of edges incident
      with a vertex, \(v\) say, we will have an even number of
      \(\sigma(v)\)'s in the product so they contribute \(+1\). If we
      have chosen an odd number of vertices we can find a smallest (in
      some arbitrarily order) such odd vertex \(v\) and we see that if
      we change the state \(\sigma\) to the state \(\sigma'\) with
      \(\sigma(v) = -\sigma'(v) \) and all other values equal, we will
      get a bijection between states witch contribute \(+1\) and
      \(-1\) and with at least one vertex of odd degree. Our
      conclusion is that we only count the choices where we have an
      even degree at each vertex. We will however count subgraphs
      where we have the opportunity to choose each edge a multiple
      number of times. If we reduce the multiple edges modulo 2 we get
      a simple subgraph of even degree. The ``surviving'' edges are
      the ones that where chosen an odd number of times so an even
      number of those ``odd'' edges have to be incident at each
      vertex.

    If we now change our view and instead of adding up the \(k\)
    moments, change the order of summation, and add up along the index
    of the number of ``surviving'' odd edges \(i\) we get a simple
    connection between the simple subgraphs and subgraphs with
    multiple edges.  We can construct an even multiedge subgraph by
    first select a simple subgraph with even degree at each vertex and
    then multiply each edge an odd number of times and then select a
    number of edges not in the even subgraph to multiply an even
    number of times. So if we first choose an even subgraph with \(i\)
    edges and multiply each edge an odd number of times we get the
    generating function \(b_i \sinh^i K\), and then choose a number of
    edges outside the even subgraph and multiply these an even number
    of times, we get the generating function \(\cosh^{m-i} K\), and we
    end up with an multiedge subgraph with an even number of edges
    incident to each vertex. If we now sum over all \(i\) we get
    \begin{equation}
      \sum_i b_i \sinh^j K \cosh^{m-i} K = \cosh^m K \sum_i b_i
      \tanh^i K
    \end{equation}
    and we see each of these graphs twice and therefor this gives
    (\ref{vdW}).
  \end{proof}
\end{theorem}

To formulate the full two variable connection we need T-joins instead
of even degree subgraphs and also consider the size of the sets in the
vertex partition induced by the state. Also note that in this theorem
we have slightly changed the meaning of \(a_{ij}\) and \(b_{ij}\) to
make the proof using standard graph theoretic notation. The following
theorem can be found in e.g. \cite{biggs}:
\begin{theorem}\label{th:I2W}
  Let \(G=G(V,E)\) be a graph, \(a_{ij}\) the number of cuts
  \(\snitt{S}\) with \(|\snitt{S}| = i\) and \(|S| = j\). Let
  \(b_{ij}\) be the number of T-joins \((T,A)\) with \(|A| = i\) and
  \(|T| = j\). Then
  \begin{displaymath}
    \sum_{ij} b_{ij} x^i y^j = 2^{-|V|} \sum_{ij} a_{ij} (1-x)^i
    (1+x)^{|E|-i} (1-y)^j (1+y)^{|V|-j}
  \end{displaymath}
  
  \begin{proof}
    Fix a subset \(T \subseteq V\) of vertices and a subset \(A
    \subseteq E\) of edges from the graph \(G = G(V,E)\). Let \(S
    \subseteq V\) be another subset of vertices and \(\snitt{S}\) be
    the cut defined by the edges from \(S\) to \(\bar{S} = V \setminus
    S\). Let the weight of the vertices in \(T \cap S\) be \(-y\), the
    weight of the vertices in \(T \cap \bar{S}\) be \(y\), the weight
    of the edges from \(A\) that lies in the cut \(\snitt{S}\) be
    \(-x\) and the rest of the edges from \(A\) have weight \(x\). Let
    the total weight of \((T,A)\) with respect to the cut
    \(\snitt{S}\) be the product of the weights of the edges and
    vertices in \((T,A)\). We say that the weight is positive if the
    coefficient in front of \(x^{|A|} y^{|T|}\) is positive and
    negative otherwise. By magnitude we denote the weight without the
    sign.

    \((T,A)\) can fail to be a T-join in basically three ways. First
    the cardinality of \(T\) can be odd, secondly there can exist a
    smallest vertex (in an arbitrary order of the vertices) \(v\)
    that is incident with an odd number of edges from \(A\) and does
    not belong to \(T\), and finally there can exist a smallest vertex
    \(v\) that is incident with an even number of edges from \(A\) and
    belongs to \(T\).

    In the first case we have two cuts \(\snitt{S}\) and \([\bar{S},
    S]\) in which the magnitude of the weight will be the same but
    with opposite sign.

    In the two latter cases we have a bijection between cuts with \(v
    \in S\) and \(v \notin S\) (we simply move the vertex \(v\)
    between \(S\) and \(\bar{S}\)) that once again give the same
    magnitude and different signs of the weight. If we sum over all
    cuts the total contribution of such a choice of \((T,A)\) will
    cancel.

    If \((T,A)\) indeed is a T-join the weight will always be positive
    since we either have an even number of vertices in \(T \cap S\)
    and an even number of edges crossing the cut or an odd number of
    vertices in \(T \cap S\) and an odd number of vertices crossing
    the cut. All in all we end up with an even number of minus signs
    and thus a positive weight.

    If we now sum over all choices \((T,A)\) and \(S\) we will count
    each T-join \(2^{|V|}\) times. If we rearrange our summation
    (i.e. we first choose a cut and then go through all
    choices of \(T\) and \(A\)) we get the theorem.
  \end{proof}
\end{theorem}

To get Theorem \ref{th:vdW} we have to shift the indices and
substitute \(x\) for \(e^{-2K}\) and \(y\) for 0 and finally double
all values since we count all cuts twice in (\ref{vdW}).
\begin{multline}
  2 e^{mK} \sum_{i,j} a_{ij} e^{-2iK}0^j = [\text{since \(0^0 = 1\)} ]
  = \sum_i 2a_i e^{(m-2i)K} = \\
  2 e^{mK} 2^{-m} \sum_{i,j} b_{ij} (1-e^{-2K})^i
  (1+e^{-2K})^{m-i} 1^n = \\
  2 \left( \frac{e^K+e^{-K}}{2} \right)^m \sum_i b_i \left(
    \frac{1-e^{-2K}}{1+e^{-2K}} \right)^i = \\
  2 \cosh^m K \sum_i b_i \tanh^i K
\end{multline}
where \(b_i\) denotes the number of T-joins with \(i\) edges. Now,
since \(a_i\) counts the number of cuts with \(i\) edges in Theorem
\ref{th:I2W}, \(2a_i e^{(m-2i)K}\) corresponds to the \(a_i e^{iK}\)
in Theorem \ref{th:vdW} via reindexing and the fact that each cut is
counted twice in Theorem \ref{th:vdW}.

Since we have a symmetry between T-joins and cuts we have the
following corollary:
\begin{corollary}
  With the same notation as in theorem \ref{th:I2W} we have
  \begin{displaymath}
    \sum_{i,j} a_{ij} x^i y^j = 2^{-|E|} \sum_{i,j} b_{ij} (1-x)^i
    (1+x)^{|V|-i} (1-y)^j (1+y)^{|E|-j}
  \end{displaymath}
  \begin{proof}
    If we choose a T-join instead of a cut the weight of \((T,A)\)
    will always be positive if and only if \((T,A)\) is a cut. In
    other cases the contributions once again cancel out.
  \end{proof}
\end{corollary}


\subsection{The thermodynamic limit}
In Physics we are interested in the so called \emph{thermodynamic
  limit} of a sequence of graphs \(\{G_j\}_{j=1}^\infty\). This is
defined as
\begin{equation}
  f(x) = \lim_{j \rightarrow \infty} \tfrac{1}{|V(G_j)|} \log Z(G_j; x, 1)
\end{equation}
when it exists. An example of such a family is \(\{C_n \times
C_n\}_{n=3}^\infty\). If the graph family is locally vertex transitive
its easy to see that the number of connected T-joins of fixed size
will grow proportionally to \(|V(G)|\). From that follows that the
total number of T-joins of a fixed size will grow as a polynomial with
degree equal to the maximal number of connected components in the
T-joins of that size and thus will the thermodynamic limit exist.

If we change notation so that our index set instead is the number of
vertices \(n\) in our graph sequence and use \(b_{j}(n)\) to denote
the number of T-joins with \(j\) edges we see that the thermodynamic
limit is
\begin{equation}
  \lim_{n \rightarrow \infty} \tfrac{1}{n} \log 2 \cosh^m x \sum_{j}
  b_{j}(n) \tanh^j x = \log 2 + d \cosh x + \sum_j b_j \tanh^j x
\end{equation}
which defines a new set of \(b_j\):s that happens to be rational
numbers and \(d = m/n\), the average degree.

\subsection{Taylor series}
It can be of interest to plot these functions, or at least some
approximation of them. Since we have a phase transition in all
interesting cases its hard to find one function that works for the
entire interval. In this section we will instead develop two Taylor
approximations, one for the high-temperature case and one for the
low-temperature case. The first one is simple, take
\begin{equation}
  u(K) = \sum_i a_i K^i = \log 2 + d \log \cosh K + \lim_{n \rightarrow
    \infty} \tfrac{1}{n} \log \sum_i b_i(n) \tanh^i K
\end{equation}
and Taylor expand around \(K=0\). If you want the function \(K(u)\) you
can invert \(u(K)\) as a Taylor series. Its often better to plot a
Pad\'e-approximation of \(u(K)\). The second one needs a little more
work:
\begin{multline}
  \frac{\dd}{\dd K} u(K) = \frac{\dd}{\dd K} \left[ \log 2 + d \log \cosh K + \lim_{n
      \rightarrow \infty} \tfrac{1}{n} \log \sum_i b_i(n) \tanh^i K
  \right] = \\
  \left[ \begin{array}{rl} x =& \tanh K \\ \tfrac{\dd}{\dd K} =&
      (1-x^2) \tfrac{\dd}{\dd x} \end{array} \right] = dx + (1-x^2)
  \frac{\dd}{\dd x} \lim_{n \rightarrow \infty} \tfrac{1}{n} \log
  \sum_i b_i(n) x^i = \\
  dx + (1-x^2) \frac{\dd}{\dd x} \lim_{n \rightarrow \infty}
  \tfrac{1}{n} \log \tfrac{1}{2^n} \sum_i a_{m-2i}(n)(1-x)^i(1+x)^{m-i}
  = \\
  dx + (1-x^2) \frac{\dd}{\dd x} \lim_{n \rightarrow \infty} \left[
    -\log 2 + d \log(1+x) + \tfrac{1}{n} \log \exp \left( n
    \sum_i a_i \left( \frac{1-x}{1+x} \right)^i \right) \right] = \\
  dx + \frac{d(1-x^2)}{1+x} + (1-x^2) \frac{\dd}{\dd x} \sum a_i
  \left( \frac{1-x}{1+x} \right)^i =
  d + (1-x^2) \frac{\dd}{\dd x} \sum a_i \left( \frac{1-x}{1+x}
  \right)^i = \\
  d + \frac{2(x-1)}{(x+1)} \sum a_i \left( \frac{1-x}{1+x}
  \right)^{i-1} =  d - 2 \sum a_i \left( \frac{1-x}{1+x} \right)^i
  = \\
  d - 2 \sum a_i \left( \frac{1 - \tanh K}{1 + \tanh K} \right)^i
\end{multline}
By not doing the last substitution \(x = \tanh K\) its easy to invert
this function and plot it in the low temperature (high energy) portion
of the scale.


\section{Two other series expansions}
From the basic thermodynamic limit one can construct two other
sequences that are of a more combinatorial flavour. These are
sequences with integer coefficients. This is because they, instead of
weighting the counts, make an explicit order in which you have to
choose things and thus avoid dividing with large factorials. The
correspondence with the thermodynamic limit series are:
\begin{equation}\label{lim}
	\sum_{i \ge 0} b_i(n) x^i = 
	\exp \left( n \sum_{i \ge 1} b_i x^i \right) =
	\prod_{i \ge 1} \left( \frac{1}{1-x^i} \right)^{n\beta_i} =
	\prod_{i \ge 1} \left( \frac{1}{1-\widehat{\beta}_i x^i} \right)^n
\end{equation}
The coefficients are connected trough these simple triangular linear equations:
\begin{align}
	i b_i &= \sum_{t|i} t \beta_t \\
	i b_i &= \sum_{rs=i} r \widehat{\beta}^s_r \\
	i \beta_i &= \sum_{rst=i} \mu(r) t \widehat{\beta}^s_t
\end{align}




\subsection{What they are counting}
The new series can be seen as placing connected subgraphs at each
vertex in some order. First we place all subgraphs (including the
empty one) rooted at vertex one, then the subgraphs rooted at vertex
two and so on in such a way that we never form a new connected
component. In this way we avoid symmetries and we get a integer
sequence. We can still calculate, in principle, the different
\(\beta_i\) and \(\widehat{\beta}_i\) for locally vertex transitive
graphs.




\section{Plots}
We shall now compare these Taylor expansions of the series with some
sampled data to compare how well behaved the series are around the
critical energy. As we shall see, the series are rather far from what
can be expected to be the truth.

\subsection{Simple cubic lattice}

\begin{figure}[!ht]
    \includegraphics[width=\textwidth]{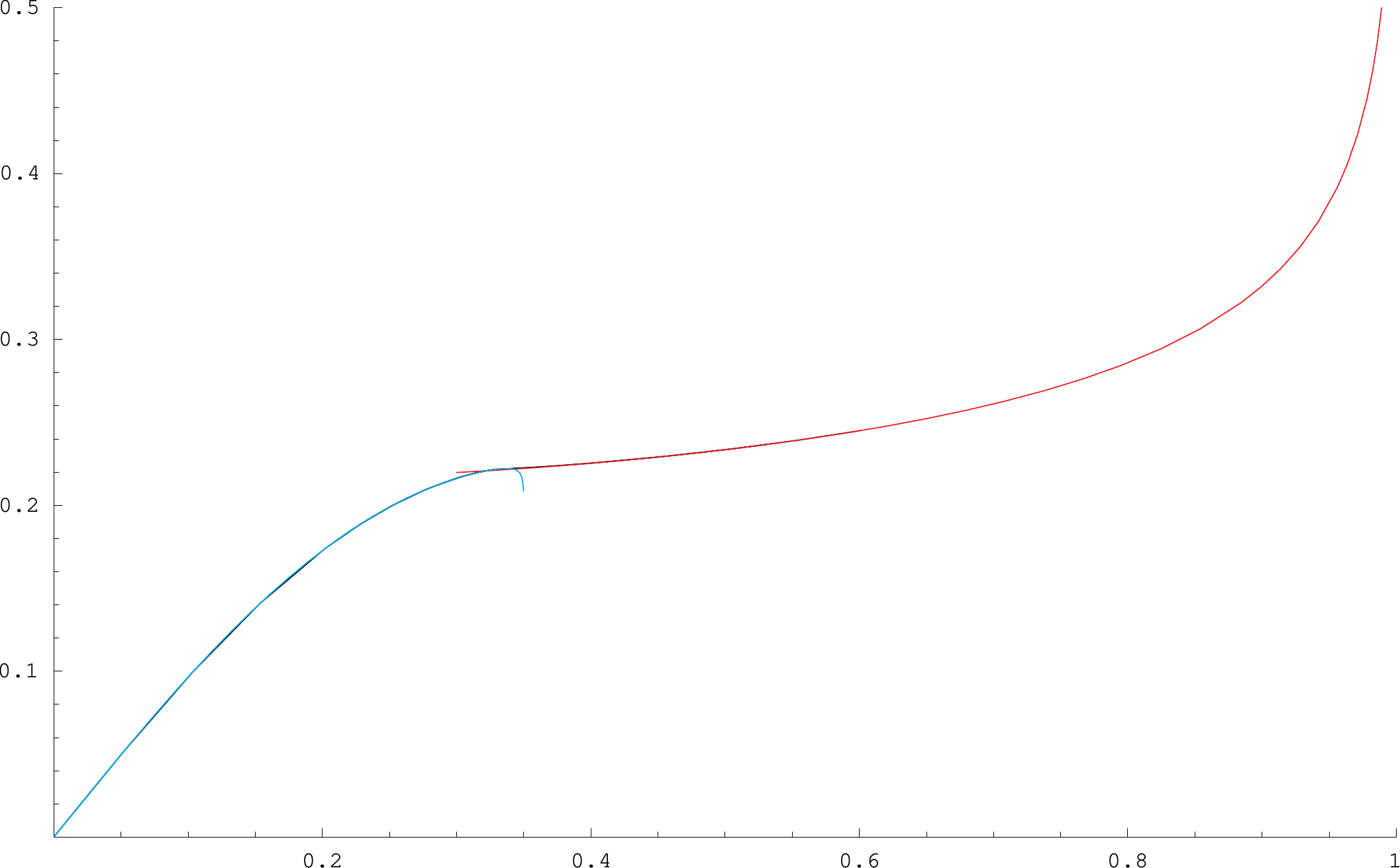}
    \caption{\(K(u)\) for the three-dimensional simple cubic lattice
      together with sampled data for the cubes of linear order 32, 64
      and 128.}\label{K:3dsc}
\end{figure}
\begin{figure}[!ht]
    \includegraphics[width=\textwidth]{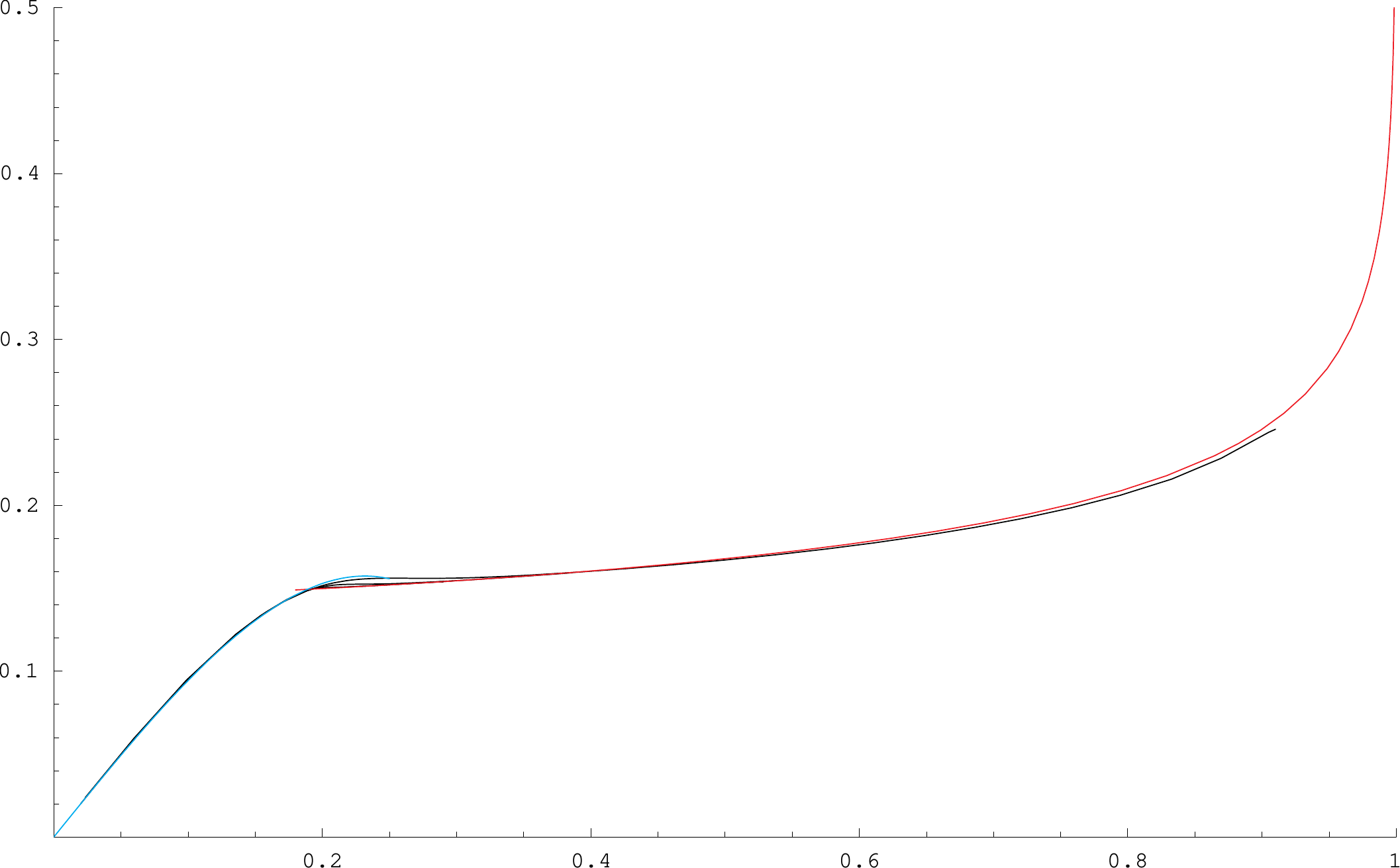}
    \caption{\(K(u)\) for the four-dimensional simple cubic lattice
      together with sampled data for the cube of linear order
      4, 6, 8, 12, 16, 32, 48 and 64.}\label{K:4dsc}
\end{figure}
\begin{figure}[!ht]
    \includegraphics[width=\textwidth]{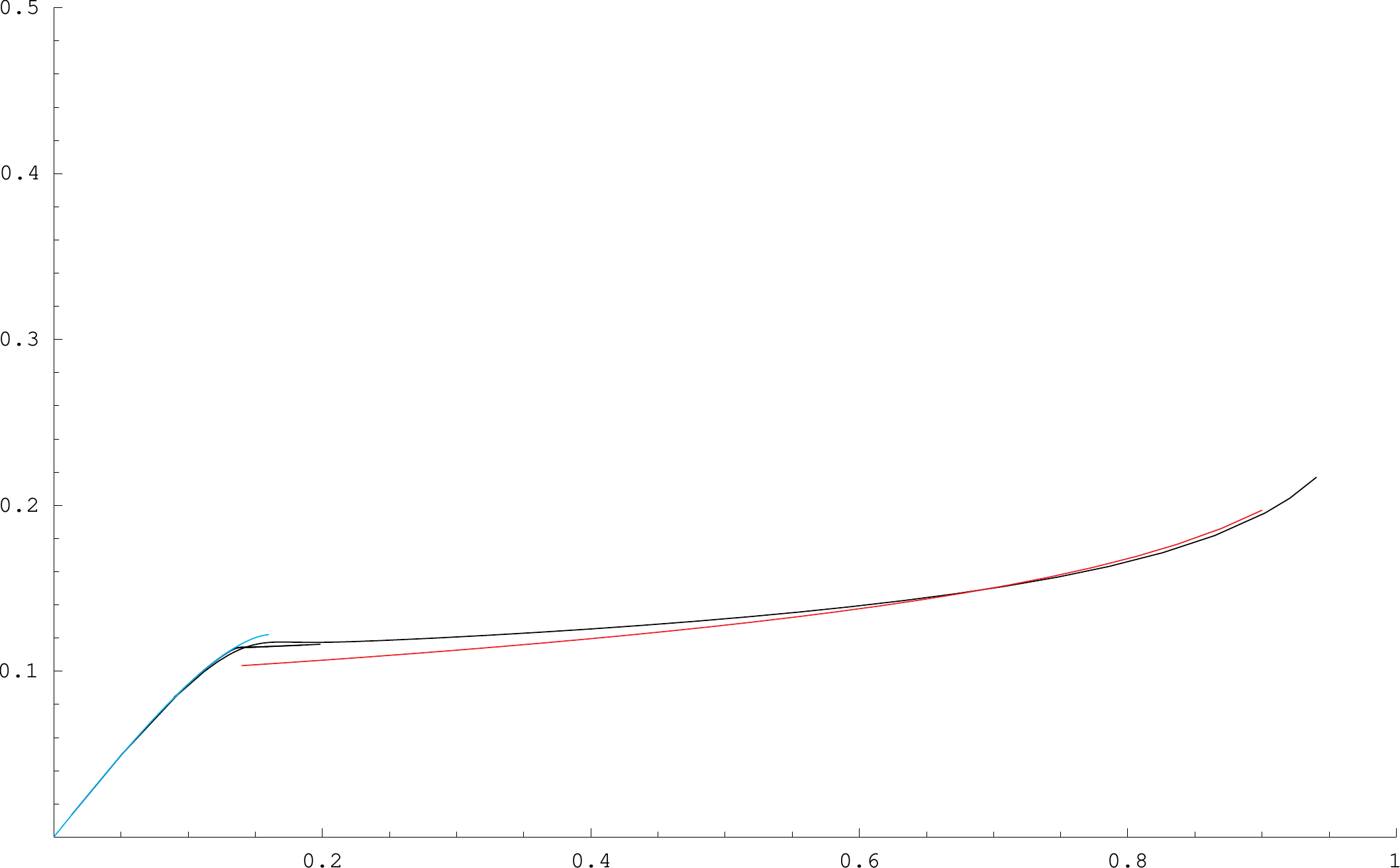}
    \caption{\(K(u)\) for the five-dimensional simple cubic lattice
      together with sampled data for the cube of linear order
      4, 6, 8, 12, 16 and 32.}\label{K:5dsc}
\end{figure}

We start of with the simple cubic lattice in three, four and five
dimensions. As can be expected, the longest series expansion is for
the three-dimensional simple cubic lattice. The four- and
five-dimensional simple cubic lattices have much shorter high- and
low-temperature expansions. Figure \ref{K:3dsc} to \ref{K:5dsc} shows
\(K(u)\) curves for the three lattices respectively. The pictures are
of diagonal Pad\'e-approximants of the Taylor expansions. The
Pad\'e-approximant for the function of the three-dimensional simple
cubic lattice is fairly accurate to about \(K \approx .22\). For the
four- and five-dimensional lattices the accuracy is a lot lower as can
be seen from the pictures.

\section{Guessing the radius of convergence} The interesting part is
of course to try to find the radius of convergence for the series of
the function \(u(K)\) since that would give the critical temperature
\(K_c\). A fairly good guess for \(K_c\) for the simple cubic lattice
is that it is approximately \(0.2216546\) (see e.g. \cite{kub}). A
simple observation is that the radius of convergence must be smaller
than any \(1/\widehat{\beta_i}^{1/i}\) since otherwise the infinite
product (\ref{lim}) will not converge. In Table \ref{tests} the row
labelled \(\min\) is the minimum over all known values of
\(\widehat{\beta_i}\) for the simple cubic lattice. In trying to
extrapolate this value we have used a polynomial of degree 3 and a min
square approximation to the data set \((1/i,
1/\widehat{\beta_i}^{1/i})\) and looked at the constant term in the
resulting polynomial. This gives the result in the row labelled
\(f(0)\). Since the coefficients is some form of combinatorial
quantity it is not unthinkable that they grow exponentially. We have
thus also tried to fit the data \((i,\widehat{\beta_i})\) to the
functions \(\alpha^i i^\beta \gamma\) and \(\alpha^i \gamma\) and
calculated \(1/\alpha\) that gives us the guessed radius of
convergence. This is the last two lines of Table \ref{tests}. The
second column is the same thing for the coefficients of \(b_i\). When
we come to the low temperature part we get into some trouble since the
coefficients of the sequences \(a_i\) and \(\widehat{\alpha_i}\) have
both positive and negative signs. If we try to use the same analysis
as for the (all positive) \(b_i\) and \(\widehat{\beta_i}\), we get
very erratic numbers, so instead we split the sequences in a positive
and a negative part and do the analysis separately. As observed by
others, the low temperature series seems to have a complex root that
is closer to the origin than the physically important real
root. Fortunately does the extrapolation with \(\alpha^i i^\beta
\gamma\) for the \(b_i\) and \(\widehat{\beta_i}\) series give a
decent idea of what the critical temperature may be.

\begin{table}
  \begin{tabular}{l|llllll}
    & \(\widehat{\beta_i}\) & \(b_i\) & \(\widehat{\alpha_i}^+\) &
    \(\widehat{\alpha_i}^-\) & \(a_i^+\) & \(a_i^-\) \\
    \hline
    \(\min\)             & 0.291989 & 0.291989 & 0.618531 & 0.616299 &
    0.618531 & 0.616307 \\
    \(f(0)\)             & 0.227727 & 0.227839 & 0.54278 & 0.543852 &
    0.542846 & 0.539196 \\
    \(\alpha^i i^\beta \gamma\) & 0.221418 & 0.221451 & 0.523726 &
    0.520324 & 0.523747 & 0.542904 \\
    \(\alpha^i \gamma\)  & 0.258385 & 0.258429 & 0.578131 & 0.565115 &
    0.578137 & 0.56453 \\
  \end{tabular}
  \caption{Different ways to try to guess the radius of
    convergence for the simple cubic lattice.} \label{tests}
\end{table}




\bibliography{taylor} 

\newcommand{\etalchar}[1]{$^{#1}$}
\begin{thebibliography}{HRL{\etalchar{+}}}

\bibitem[Big77]{biggs}
Norman Biggs.
\newblock {\em Interaction models}.
\newblock Cambridge University Press, Cambridge, 1977.
\newblock Course given at Royal Holloway College, University of London,
  October--December 1976, London Mathematical Society Lecture Note Series, No.
  30.

\bibitem[Cip87]{cipra}
Barry~A. Cipra.
\newblock An introduction to the {I}sing model.
\newblock {\em Amer. Math. Monthly}, 94(10):937--959, 1987.

\bibitem[HRL{\etalchar{+}}]{kub}
Roland H\"aggkvist, Andres Rosengren, Per~H\r{a}kan Lundow, Klas Markstr\"om,
  Daniel Andr\'en, and Petras Kundrotas.
\newblock On the {I}sing model for the simple cubic lattice.
\newblock Manuscript.

\bibitem[Isi25]{ising}
Ernst Ising.
\newblock Beitrag zur {T}heorie des {F}erromagnetismus.
\newblock {\em Z.Physik}, 31:253--258, 1925.

\bibitem[Len20]{lenz}
Wilhelm Lenz.
\newblock Beitrag zum {V}erst{\"a}ndnis der magnetishen {E}rscheinungen in
  festen {K}{\"o}rpern.
\newblock {\em Z. Physik}, 21:613--615, 1920.

\bibitem[Ons44]{onsager}
Lars Onsager.
\newblock Crystal statistics. {I}. {A} two-dimensional model with an
  order-disorder transition.
\newblock {\em Phys. Rev. (2)}, 65:117--149, 1944.

\end{thebibliography}
\bibliographystyle{alpha}


\newpage
\appendix




\section{Tables}
We have collected the various series we have found in this
appendix. Some of these are old and thus not especially long and some
are from newer calculations and longer.

\begin{table}[ht]
  \center
  \begin{tabular}{r|rrr}
    \(n\) & \(\alpha_n\) & \(\widehat{\alpha}_n\) & \(a_n\) \\
    \hline
    6 & 1 & 1 & 1 \\
10 & 3 & 3 & 3 \\
12 & -4 & -4 & -7/2 \\
14 & 15 & 15 & 15 \\
16 & -33 & -33 & -33 \\
18 & 104 & 104 & 313/3 \\
20 & -282 & -285 & -561/2 \\
22 & 849 & 849 & 849 \\
24 & -2460 & -2470 & -9847/4 \\
26 & 7485 & 7485 & 7485 \\
28 & -22542 & -22647 & -45069/2 \\
30 & 69392 & 69384 & 346966/5 \\
32 & -213738 & -214299 & -427509/2 \\
34 & 666750 & 666750 & 666750 \\
36 & -2086785 & -2092121 & -12520405/6 \\
38 & 6583341 & 6583341 & 6583341 \\
40 & -20852223 & -20892996 & -83409453/4 \\
42 & 66425750 & 66424630 & 464980286/7 \\
44 & -212410377 & -212770353 & -424819905/2 \\
46 & 682202205 & 682202205 & 682202205 \\
48 & -2198562644 & -2201602421 & -17588511087/8 \\
50 & 7110521070 & 7110521022 & 35552605353/5 \\
52 & -23065955826 & -23093964696 & -46131904167/2 \\
54 & 75045653088 & 75045278168 & 675410878105/9 \\
56 & -244806881325 & -245063348553 & -979227570369/4 \\
58 & 800606679471 & 800606679471 & 800606679471 \\
60 & -2624325216574 & -2626724535242 & -13121625909861/5 \\
62 & 8621219166681 & 8621219166681 & 8621219166681 \\
64 & -28379404026078 & -28402366460136 & -113517616531821/4 \\

  \end{tabular}
  \caption{Simple cubic lattice}
\end{table}

\begin{sidewaystable}[ht]
  \center
  \begin{tabular}{r|rrr}
    \(n\) & \(\beta_n\) & \(\widehat{\beta}_n\) & \(b_n\) \\
    \hline
    4 & 3 & 3 & 3 \\
6 & 22 & 22 & 22 \\
8 & 186 & 183 & 375/2 \\
10 & 1980 & 1980 & 1980 \\
12 & 24032 & 23793 & 24044 \\
14 & 319170 & 319170 & 319170 \\
16 & 4514664 & 4497993 & 18059031/4 \\
18 & 67003462 & 66999920 & 201010408/3 \\
20 & 1032455736 & 1030496478 & 5162283633/5 \\
22 & 16397040750 & 16397040750 & 16397040750 \\
24 & 266958785298 & 266673642443 & 266958797382 \\
26 & 4437596650548 & 4437596650548 & 4437596650548 \\
28 & 75078511535604 & 75027576950427 & 525549581866326/7 \\
30 & 1289656872697576 & 1289654284203514 & 6448284363491202/5 \\
32 & 22447149807206352 & 22437033558570606 & 179577198475709847/8 \\
34 & 395251648062268272 & 395251648062268272 & 395251648062268272 \\
36 & 7031220729573330428 & 7028971745154518717 & 21093662188820520521/3 \\
38 & 126225408651399082182 & 126225408651399082182 & 126225408651399082182 \\
40 & 2284608766597864770492 & 2284077801219364370517 & 4569217533196761997785/2 \\
42 & 41655898158709803301884 & 41655887320814523000436 & 291591287110968623857940/7 \\
44 & 764611114761740442269316 & 764476683289070060492337 & 8410722262379235048686604/11 \\
46 & 14120314204713719766888210 & 14120314204713719766888210 & 14120314204713719766888210 \\

  \end{tabular}
  \caption{Simple cubic lattice}
\end{sidewaystable}

\begin{table}[ht]
  \center
  \begin{tabular}{r|rrrr}
    \(n\) & \(a^4_n\) & \(b^4_n\) & \(a^5_n\) & \(b^5_n\) \\
    \hline
    4 & 0 & 6 & 0 & 10 \\
6 & 0 & 76 & 0 & 180 \\
8 & 1 & 1371 & 0 & 5025 \\
10 & 0 & 30152 & 1 & 178696 \\
18 & 0 &  & 5 &  \\
20 & 28 &  & -11/2 &  \\
22 & -64 &  & 45 &  \\
24 & 127/3 &  & -95 &  \\
26 & 228 &  & 50 &  \\
28 & -834 &  & 5 &  \\
30 & 1116 &  & 1471/3 &  \\
32 & 5911/4 &  & -1685 &  \\
34 & -10404 &  & 1885 &  \\
36 & 21460 &  & -775/2 &  \\
38 & -1956 &  & 5445 &  \\
40 & -595179/5 &  & -112951/4 &  \\
42 & 1076092/3 &  & 49545 &  \\
44 & -344316 &  & -62795/2 &  \\
46 & -1132588 &  & 71442 &  \\
48 & 10842287/2 &  &  &  \\
50 & -9187444 &  &  &  \\
52 & -5820150 &  &  &  \\
54 & 73867260 &  &  &  \\
56 & -1294335811/7 &  &  &  \\
58 & 95069292 &  &  &  \\
60 & 2609680726/3 &  &  &  \\
62 & -3217644924 &  &  &  \\

  \end{tabular}
  \caption{Simple cubic lattice in dimension 4 (\(x^4_n\)) and 5 (\(x^5_n\))}
\end{table}




\begin{table}[ht]
  \center
  \begin{tabular}{r|rr}
    \(n\) & \(a_n\) & \(b_n\) \\
    \hline
    4 & 0 & 6 \\
6 & 2 & 44 \\
7 & 0 & 36 \\
8 & 0 & 384 \\
9 & 0 & 688 \\
10 & 6 & 4572 \\
11 & 6 & 11148 \\
12 & -14 & 66158 \\
13 & 0 & 190662 \\
14 & 30 & 1051668 \\
15 & 60 & 3471452 \\
16 & -108 & 17917704 \\
17 & -144 & 65438160 \\
18 & 1118/3 & 971627006/3 \\
19 & 498 & 1265584728 \\
20 & -714 & 30625029636/5 \\
21 & -2366 & 25078631014 \\
22 & 3270 & -3816568476 \\
23 & 7704 &  \\
24 & -8106 &  \\
25 & -27372 &  \\
26 & 19842 &  \\
27 & 114342 &  \\
28 & -68892 &  \\
29 & -377376 &  \\
30 & 643952/5 &  \\
31 & 1431726 &  \\
32 & -137718 &  \\
33 & -5365756 &  \\
34 & -33330 &  \\
35 & 18644574 &  \\
36 & -970922/3 &  \\
37 & -69012330 &  \\
38 & -32516754 &  \\
39 & 249820316 &  \\
40 & 162829320 &  \\
41 & -869879742 &  \\
42 & -5660822830/7 &  \\
43 & 3155460756 &  \\

  \end{tabular}
  \caption{Simple cubic lattice, 3-states Potts model}
\end{table}


\end{document}